\documentstyle[12pt]{article}
\def\overlay#1#2{\setbox0=\hbox{#1}\setbox1=\hbox to \wd0{\hss
#2\hss}#1\hskip -2\wd0\copy1}
\def\lsim{\mathrel{\rlap{\lower4pt\hbox{\hskip1pt$\sim$}}
    \raise1pt\hbox{$<$}}}         %less than or approx. symbol
\def\gsim{\mathrel{\rlap{\lower4pt\hbox{\hskip1pt$\sim$}}
    \raise1pt\hbox{$>$}}}         %greater than or approx. symbol
\def\beq{\begin{equation}}
\def\eeq{\end{equation}}
\def\bea{\begin{eqnarray}}
\def\eea{\end{eqnarray}}

\begin{document}
\begin{center}
{\large \bf q-Deformed Boson Expansions}
\end{center}
\vskip 2.0mm
\begin{center}
S.S. Avancini$^1$,F.F. de Souza Cruz$^{1,2}$, J.R. Marinelli$^1$, D.P. Menezes$^1$ and  M.M.
Watanabe de Moraes$^1$
\end{center}
\vskip 2.0mm
\begin{center}
$^1${\it Departamento de F\'{\i}sica, Universidade Federal 
de Santa Catarina,\\ 
88.040-900 Florian\'opolis - S.C., Brazil\\ }
$^2${\it Institute for Nuclear Theory, University of Washington,\\
Seattle ,WA 98195,USA\\}
\end{center}

\vspace{2.5cm}

\begin{abstract}
A deformed boson mapping of the Marumori type is derived for 
an underlying $su(2)$ algebra. As an example, we bosonize a
pairing hamiltonian in a two level space, for which an exact
treatment is possible. Comparisons are then made between the
exact result, our q- deformed boson expansion and the usual 
non - deformed expansion.
\end{abstract}
%
%\pacs{PACS number(s) }
\newpage

Nowadays increasing importance has been given to {\it quantum 
algebraic}
applications in several fields of physics \cite{qa}. In many cases, 
when the
usual Lie algebras do not suffice to explain certain physical behaviors,
quantum algebras are found to be successful mainly due to a
free deformation parameter. In these cases, it is expected that a
physical meaning be attached to the deformation parameter, but this
is still a very challenging question. For an extensive review
article on the subject, refer to \cite{bon}.
In this work we are concerned with possible improvements that
quantum algebras may add to boson expansions (or boson mappings).

In the literature it is easy to find situations in which fermion
pairs can be replaced by bosons. This is normally performed 
with the help of boson mappings, that link the fermionic Hilbert
space to another Hilbert space constructed with bosons. Of course
boson mapping techniques are only useful when the Pauli Principle
effects are somehow minimized. Historically boson expansion theories
were introduced from two different points of view. The first one
is the Beliaev - Zelevinsky - Marshalek (BZM) method \cite{bzm}, which 
focuses on the mapping of operators by requiring that the boson images
satisfy the same commutation relations as the fermion operators. 
In principle, all important operators can be constructed from a set of
basic operators whose commutation relations form an algebra. The 
mapping is achieved by preserving this algebra and mapping these basic 
operators. The second one is the Marumori method \cite{marumori},
which focuses on the mapping of state vectors. This method defines
the operator in such a way that the matrix elements are conserved
by the mapping and the importance of the commutation rules is
left as a consequence of the requirement that matrix elements coincide
in both spaces.
The BZM and the Marumori expansions are equivalent 
at infinite order, which means that just with the proper mathematics
one can go from one expansion to the other.
  
In this letter we concentrate on this second boson mapping method.
First of all, we briefly outline the main aspects of the mapping from
a fermionic space to a quantum deformed bosonic space. Once the 
deformation parameter is set equal to one, the usual boson expansion
is recovered. Then the simple pairing interaction model is used as
an example for our calculations. The pairing hamiltonian is
exactly diagonalized and the results are compared with the ones 
obtained from the traditional boson and from the q- deformed
boson expansions. In both cases we analyse the results for the 
second and fourth order hamiltonians.

In what follows we show a Marumori type deformed boson mapping. We
start from an arbitrary operator $\hat O$ acting on a finite 
fermionic space. This fermionic Hilbert space with dimension $N+1$
is spanned by a basis formed by the states $\{ |n> \}$, with $n=0,
1,...N$. Hence,
 
\beq
\hat O = \sum_{n,n'=0}^N <n'|\hat O|n> |n'><n| .
\label{opf}
\eeq
In order to obtain the boson operators, we map $\hat O \rightarrow
\hat O_B$ :
\beq
\hat O_B = \sum_{n,n'=0}^N <n'|\hat O|n> |n')(n| ,
\label{opb}
\eeq
 where
\beq
|n)={1\over \sqrt{[n]!}} (b^{\dag})^n |0)
\eeq
are the deformed boson states \cite{dbs}
with $[n]={q^n-1\over q-1}$ 
and $[b,b^{\dag}]_q=bb^{\dag}-qb^{\dag}b =1$. Note that the
usual brackets $<|>$ stand for fermionic states and the
round brackets $(|)$ stand for bosonic states. From the above
considerations, it is straightforward to check that
\beq
<m|\hat O|m'>=(m|\hat O_B|m').
\eeq
Therefore, we notice that the mapping is achieved by the
equality between the matrix elements in the fermionic space 
and their counterparts in the bosonic space. As examples, we 
show the expressions for the $su(2)$ operators in the deformed 
bosonic space:
\beq
(J_z)_B= \sum_{n=0}^{2j} \sum_{l=0}^{\infty} (-j+n) 
{ (-1)^l q^{l(l-1)/2} \over [n]! [l]!} (b^{\dag})^{n+l} b^{n+l},
\label{maz}
\eeq

\beq
(J_+)_B= \sum_{n=0}^{2j} \sum_{l=0}^{\infty} \sqrt{(n+1)(2j-n)\over 
[n+1]} 
{ (-1)^l q^{l(l-1)/2} \over [n]! [l]!} (b^{\dag})^{n+l+1} b^{n+l},
\label{map}
\eeq
 
\beq
(J_+ J_-)_B= \sum_{n=0}^{2j} \sum_{l=0}^{\infty} n (2j-n+1)
{ (-1)^l q^{l(l-1)/2} \over [n]! [l]!} (b^{\dag})^{n+l} b^{n+l},
\label{pm}
\eeq

\beq
(J_- J_+)_B= \sum_{n=0}^{2j} \sum_{l=0}^{\infty} (2j-n)(n+1)
{ (-1)^l q^{l(l-1)/2} \over [n]! [l]!} (b^{\dag})^{n+l} b^{n+l},
\label{mp}
\eeq
and $(J_-)_B=(J_+)_B^{\dag}$. In deducing the above expressions
we have used that \cite{fiv}

\beq
|0><0|=:exp_q(-b^{\dag}b):=\sum_{l=0}^{\infty}
{ (-1)^l q^{l(l-1)/2} \over [l]!} (b^{\dag})^l b^l,
\label{vacuo}
\eeq

and we define the $su(2)$ basis as usual, i.e., $|n>=|j m>$, with
$m=-j+n$.

Next, we apply the q- deformed boson expansions to the
pairing interaction model \cite{krieger}, which
consists of two N-fold degenerate levels, whose energy
difference is $\epsilon$. The lower level has energy $-\epsilon/2$ 
and its single-particle states are usually labelled $j_1 m_1$ and 
the upper
level has energy $\epsilon/2$ and its single-particle states are
labelled $j_2 m_2$. The pairing hamiltonian reads \cite{cambia}:

\beq
H={\epsilon \over 2} \sum_m (a^{\dag}_{j_1 m} a_{j_1 m} -
  a^{\dag}_{j_2 m} a_{j_2 m}) 
-{G \over 4} \left(
\sum_j \sum_{m} a^{\dag}_{j m} a^{\dag}_{j \bar m} 
\sum_{j'} \sum_{m'} a_{j' \bar m'} a_{j'm'} + h.c. \right)
\label{hpair1}
\eeq
where $a^{\dag}_{j \bar m}= (-1)^{j-m} a_{j -m}$ . In what follows, 
the number of particles (which are fermions) $N$ will be even and 
$2j=N/2$.
Introducing the quasispin $su(2)$ generators :
$$S_+=S_-^{\dag}={1 \over2} \sum_{m_1} a^{\dag}_{j_1 m_1} 
a^{\dag}_{j_1 \bar m_1} = \sqrt{\Omega} A^{\dag}_1$$
$$S_z={1\over 2} \sum_{m_1} a^{\dag}_{j_1 m_1} a_{j_1 m_1} - 
{N \over 4}$$
$$L_+=L_-^{\dag}={1\over2} \sum_{m_2} a^{\dag}_{j_2 m_2} 
a^{\dag}_{j_2 \bar m_2} = \sqrt{\Omega} A^{\dag}_2$$
$$L_z={1\over 2} \sum_{m_2} a^{\dag}_{j_2 m_2} a_{j_2 m_2} - 
{N \over 4}$$
one sees that the pairing interaction has an underlying  
$su(2) \otimes su(2)$ algebra. With the help of these operators,
eq. (\ref{hpair1}) can be rewritten as

\beq
H=\epsilon(S_z-L_z)-{G \Omega \over 2}
\left( (A_1^{\dag}+A_2^{\dag})(A_1+A_2)+ 
(A_1+A_2)(A_1^{\dag}+A_2^{\dag})
\right).
\label{hpair2}
\eeq

 The basis of states used for the diagonalization of the above 
hamiltonian is
$|S={N\over4}~~L_z~,~L={N\over4}~~-L_z>$ \cite{krieger}, \cite{ours}. 

Deformation can be straightforwardly introduced by deforming the 
$su(2) \otimes su(2)$ algebra and this problem has already been
tackled in ref. \cite{ours}. To check the validity of the boson
expansion method proposed in this letter, we substitute eqs. 
(\ref{maz}), (\ref{map}),
(\ref{pm}) and (\ref{mp}) into eq. (\ref{hpair2})
and obtain for the fourth order hamiltonian:

$${H_4 \over \epsilon} = -{x \over 2} + 
\left(1-{x (\Omega-1) \over 2 \Omega}\right) b^{\dag}_1 b_1 +
\left(-1-{x (\Omega-1) \over 2 \Omega}\right)b^{\dag}_2 b_2 
-{x \over2} (b^{\dag}_1 b_2 + b^{\dag}_2 b_1)$$
$$+\left({2 \over [2]}-1\right)(b^{\dag}_1 b^{\dag}_1 b_1 b_1
- b^{\dag}_2 b^{\dag}_2 b_2 b_2)$$
$${-x \over 4 \Omega} \left(2 - 3 \Omega -{8 \over [2]} +
{5 \Omega \over [2]} + {\Omega \over [2]}q \right)
(b^{\dag}_1 b^{\dag}_1 b_1 b_1
+ b^{\dag}_2 b^{\dag}_2 b_2 b_2)$$
\beq
-{x \over 2 \Omega}\left(\sqrt{2 \Omega(\Omega-1) \over [2]}
- \Omega \right) (b^{\dag}_1 b^{\dag}_2 b_2 b_2
+ b^{\dag}_1 b^{\dag}_1 b_1 b_2 + h.c.)
\label{h4}
\eeq
where $x=2 G \Omega / \epsilon$. The second order hamiltonian 
is easily read off from the above equation by omitting all
terms containing four boson operators. Diagonalizing eq. 
(\ref{h4}) is a simple task and for this purpose the basis
used is
\beq
|n_1 n_2>= {1 \over \sqrt{ [n_1]! [n_2]!}}
(b^{\dag}_1)^{n_1} (b^{\dag}_2)^{n_2} |0>
\label{basis}
\eeq
and  
$$b^{\dag}_1 |n_1> = \sqrt{[n_1 +1]} |n_1+1>~~~,~~~
b_1 |n_1> = \sqrt{[n_1]} |n_1-1>$$
with similar expressions for the $b_2$ and $b^{\dag}_2$ operators.
We finally obtain:
$${H_4 \over \epsilon} |n_1 n_2> = \left(
-{x \over 2} + 
\left({2 \over [2]}-1\right)([n_1][n_1-1] - [n_2][n_2-1]) \right.$$
$$+ \left(1-{x (\Omega-1) \over 2 \Omega}\right) [n_1] +
\left(-1-{x (\Omega-1) \over 2 \Omega}\right) [n_2]$$ 
$$\left.{-x \over 4 \Omega} \left(2 - 3 \Omega -{8 \over [2]} +
{5 \Omega \over [2]} + {\Omega \over [2]}q\right)
([n_1][n_1-1] + [n_2][n_2-1]) \right)
|n_1 n_2>$$
$$+(-{x \over2} \sqrt{[n_2][n_1+1]} -
{x \over 2 \Omega} ( \sqrt{2 \Omega(\Omega-1) \over [2]}
- \Omega ) ([n_2-1]+[n_1]) \sqrt{[n_2][n_1+1]}) |n_1+1~ n_2-1>$$
\beq
+(-{x \over2} \sqrt{[n_1][n_2+1]} -
{x \over 2 \Omega} ( \sqrt{2 \Omega(\Omega-1) \over [2]}
- \Omega ) ([n_1-1]+[n_2]) \sqrt{[n_2+1][n_1]}) |n_1-1~ n_2+1>
\label{hfinal}
\eeq

Eq. (\ref{hfinal}) yields the energy spectrum for the deformed
Marumori type boson expansion. When $q$ is set equal to unity,
the non- deformed spectrum is obtained. In what follows,
we have chosen $x=1.0$
and the degeneracy $\Omega=20$. In figure 1 we show the
ground state energy resulting from the exact diagonalization of
eq. (\ref{hpair2}) and the ground state energies obtained from
the second and fourth order hamiltonians defined in eq. (\ref{h4}) as
a function of the number of pairs for $q=1$. One can see that the
fourth order curve lies closer to the exact result than the second
order curve, as expected, once the full expansion converges to
 the exact result.

We then compare the exact result with the deformed second and fourth
order expansions and the results are plotted in figure 2. Setting
$q=0.862$, we find that the second order expansion converges to the 
exact result and for $q=0.810$ the fourth order expansion also
converges. This implies that the deformation parameter is playing the
same r\^ole as all the rest of the truncated expansion. One does
not have to go beyond the deformed second order boson expansion
to obtain the exact result while the fourth order non- deformed
expansion gives still very poor results, as seen in figure 1. 
Therefore, the use of quantum algebras in boson expansion theories
can be a very useful method in providing the same result as 
the complete series. At this respect, we believe that further
investigations, like the consideration of the BZM method and also
of other model hamiltonians, deserve some effort in the future.

\vskip 0.45in

This work has been partially supported by CNPq .

\newpage
\section{Figure Captions}

Figure 1) The ground state energy $E_0$ is plotted as a function
of the number of pairs for the exact result (solid line),
the second order expansion result (short- dashed line) and
for the fourth order result (long- dashed line) for $q=1$,
the interaction strenght $x=1.0$ and the degeneracy $\Omega=20$.

Figure 2) The ground state energy $E_0$ is plotted as a function
of the number of pairs for the exact result with $q=1$ (solid line),
the second order expansion result with $q=0.862$ (dashed line) and
for the fourth order result with $q=0.810$ (dot- dashed line) for
the interaction strenght $x=1.0$ and the degeneracy $\Omega=20$.

\end{document}